\documentstyle[prb,aps,twocolumn]{revtex}  
\input epsf
\epsfverbosetrue
\tighten
\begin{document}
\draft
\twocolumn[\hsize\textwidth\columnwidth\hsize\csname @twocolumnfalse\endcsname

\title{Twin wall of cubic-tetragonal ferroelastics}
 
\author{S. H. Curnoe\cite{Stephanie} and A. E. Jacobs\cite{Allan}}
\address{Department of Physics, University of Toronto\\
Toronto, Ontario, CANADA M5S 1A7}
 
\date{\today}
 
\maketitle
 
\begin{abstract}
We derive solutions for the twin wall linking two tetragonal variants of the 
cubic-tetragonal ferroelastic transformation, including for the first time 
the dilatational and shear energies and strains. 
Our solutions satisfy the compatibility relations exactly and are obtained at 
all temperatures. 
They require four non-vanishing strains except at the Barsch-Krumhansl 
temperature $T_{BK}$ (where only the two deviatoric strains are needed). 
Between the critical temperature and $T_{BK}$, material in the wall region is 
dilated, while below $T_{BK}$ it is compressed. 
In agreement with experiment and more general theory, the twin wall lies in a 
cubic 110-type plane. 
We obtain the wall energy numerically as a function of temperature and we 
derive a simple estimate which agrees well with these values. 

\end{abstract}
 
]
\narrowtext
\tightenlines


Ferroelastic transformations are diffusionless, first-order, shape-changing
phase changes in the solid state\cite{aizu69,salje93}. 
In cubic-tetragonal (C-T) systems like Nb$_3$Sn, V$_3$Si, In-Tl alloys, 
Fe-Pd alloys and Ni$_2$MnGa,
the cubic unit cell elongates (or contracts) along one of 
three axes to form a tetragonal unit cell; 
below the transition temperature $T_c$ there are three possible homogeneous 
products (variants) differing only in orientation. 

Barsch and Krumhansl\cite{bk84} (BK) obtained an analytical solution for the 
twin wall linking two tetragonal variants of C-T ferroelastics. 
The dilatational and shear strains vanish identically, and the two remaining 
strains are functions of a single coordinate. 
The wall lies in a cubic 110-type plane, in agreement with experiment 
and as already known on more general grounds\cite{sapriel75}. 

The BK solution is valid, however, at only a single temperature ($T=T_{BK}$). 
At any other $T$, the dilatational and shear strains are not zero, and the 
only known method to find the C-T twin wall structure requires solving the 
full three-dimensional (3D) partial differential equations (previous 
attempts\cite{yamada,hongolson} at a 1D solution 
omitted the non-deviatoric strains). 
Some numerical solutions\cite{kl1,kl2,rasmus} of these equations have been 
obtained, but no results have been given for the wall structure. 

That is, 16 years after BK\cite{bk84}, the wall structure is still unknown 
except at $T_{BK}$. 

The following solves this long-standing problem, which is of considerable 
physical interest given the large strains and large magnetic-field 
effects\cite{james} in Ni$_2$MnGa. 
Specifically, we present a 1D solution for the C-T twin wall at all $T$. 
Our solutions, which include dilatational and shear energies and strains, 
satisfy the compatibility relations by virtue of analytical relations that 
we derive between the strains; 
these relations allow us to reduce the problem to the solution of ordinary, 
rather than partial, differential equations. 
We recover the BK solution at $T_{BK}$ and we present results at both 
higher and lower $T$. 

The three paragraphs immediately following define the strains and the two 
parts of the free-energy density. 
The next two paragraphs obtain the key new results of our analysis, namely 
two relations between the strains, and the resulting Euler-Lagrange equations. 
The remaining paragraphs discuss the results from solution of these equations. 
We find that the twin-wall region is dilated near $T_c$, and compressed below 
$T_{BK}$. 

Six strains are required to describe C-T ferroelastics, the dilatational 
strain $e_1$, the deviatoric strains $e_2$ and $e_3$, and the shear strains 
$e_4$, $e_5$ and $e_6$. 
With the coordinate axes along the four-fold axes, and in the small-strain 
approximation, these are 
\begin{equation}
\begin{array}{ll}
e_1=\left(u_{1,1}+u_{2,2}+ u_{3,3}\right)/\sqrt{3} \ ,&
                   e_2=\left(u_{1,1}-u_{2,2}\right)/\sqrt{2}\ , \\
e_3=\left(u_{1,1}+u_{2,2}-2u_{3,3}\right)/\sqrt{6} \ ,&  
                   e_6=\left(u_{1,2}+u_{2,1}\right)/2           
\end{array}
\end{equation}
plus obvious expressions for $e_4$ and $e_5$. 
Here ${\bf u}=(u_1,u_2,u_3)$ is the displacement 
of the material point originally at ${\bf x}$, and 
$u_{i,j}=\partial_ju_i=\partial u_i/\partial x_j$. 
We need also the components $\omega_3 =(u_{1,2}-u_{2,1})/2, \ {\it etc.}$ 
of the local rotation {\boldmath$\omega$}. 

The free energy $F$ is the integral 
$F=\int_V {\cal F}\ d^3x$ 
of the free-energy density ${\cal F}$ over the undeformed volume $V$. 
For proper ferroelastics (where the strain is the primary order parameter), 
${\cal F}$ is the sum of strain and strain-gradient parts. 
The strain part is 
$${\cal F}_s= 
 \frac{A_1}{2}      e_1^2
+\frac{A_2}{2}\left(e_2^2+e_3^2\right)
-\frac{B_2}{3}\left(e_3^3-3e_2^2e_3\right) $$
\begin{equation}
+\frac{C_2}{4}\left(e_2^2+e_3^2\right)^2
+\frac{A_4}{2}\left(e_4^2+e_5^2+e_6^2\right)\ .
\end{equation}
In Voigt notation, the dilatational and shear constants are 
$A_1=C_{11}+C_{12}$ and $A_4=4C_{44}$ respectively, with both $>0$ for 
stability; the corresponding terms in the density were omitted in 
previous treatments\cite{bk84,yamada,hongolson}.
The coefficient $A_2$ depends on temperature as $A_2=A_2^\prime(T-T_0)$, 
where $T_0$ is the stability limit of the cubic phase and $A_2^\prime$ is 
a material-dependent constant; 
above $T_c$, $A_2=C_{11}-C_{12}$. 
For $A_2>{1\over4}B_2^2/C_2$, the energy has only the cubic minimum (all 
strains zero). 
For $A_2<{1\over4}B_2^2/C_2$, there are in addition
three degenerate minima symmetrically 
located in the plane of the deviatoric strains: 
\renewcommand{\theequation}{3\alph{equation}}
\setcounter{equation}{0}
\begin{eqnarray} 
e_2=0                \ ,\ e_3&=& e_{30}   \\
e_2=-\sqrt{3}e_{30}/2\ ,\ e_3&=&-e_{30}/2 \\
e_2= \sqrt{3}e_{30}/2\ ,\ e_3&=&-e_{30}/2 
\end{eqnarray} 
with $e_1=e_4=e_5=e_6=0$ and 
\renewcommand{\theequation}{\arabic{equation}}
\setcounter{equation}{3}
\begin{equation} 
e_{30}=\left[B_2+\left(B_2^2-4A_2C_2\right)^{1/2}\right]/\left(2C_2\right)\ ;
\end{equation} 
the tetragonal four-fold axes are in the 3, 1 and 2 directions respectively. 
The C-T transition, which is first-order, occurs at $A_2={2\over9}B_2^2/C_2$ 
where $e_{30}={2\over3}B_2/C_2$. 

The free-energy density requires also strain-gradient terms, so that energy 
is required to introduce variant-variant walls (otherwise the system can 
subdivide into arbitrarily fine variants). 
We keep only the two invariants quadratic in the deviatoric strain 
derivatives: 
$$
{\cal F}_{sg}=
 \frac{d_2}{2}\left[ \left(e'           _{2,1}\right)^2
                    +\left(e''          _{2,2}\right)^2
                    +\left(e\phantom{'}_{2,3}\right)^2\right] 
$$
\begin{equation}
+\frac{d_3}{2}\left[ \left(e'           _{3,1}\right)^2
                    +\left(e''          _{3,2}\right)^2
                    +\left(e\phantom{'}_{3,3}\right)^2\right]
\end{equation} 
where $e'_2$, $e''_2$, $e'_3$ and $e''_3$ are obtained from $e_2$ and $e_3$ 
by $2\pi/3$ rotations about the cubic 111 axis: 
\begin{equation} 
\begin{array}{rl}
e_2' = \left(u_{2,2}-u_{3,3}         \right)/\sqrt{2}
    =&\left(-e_2+\sqrt{3}e_3        \right)/2\ ,\\
e_2''= \left(u_{3,3}-u_{1,1}         \right)/\sqrt{2}   
    =&\left(-e_2-\sqrt{3}e_3        \right)/2\ ,\\
e_3' = \left(u_{2,2}+u_{3,3}-2u_{1,1}\right)/\sqrt{6}
    =&\left(-e_3-\sqrt{3}e_2        \right)/2\ ,\\
e_3''= \left(u_{3,3}+u_{1,1}-2u_{2,2}\right)/\sqrt{6}
    =&\left(-e_3+\sqrt{3}e_2        \right)/2\ .
\end{array}
\end{equation} 
Both terms are transparently invariant and non-negative, and so we have the 
stability requirements $d_2\geq0$ and $d_3\geq0$ (which differ from those in 
Ref.\onlinecite{hongolson}); 
contact with previous treatments\cite{bk84,hongolson,rasmus} is made by 
writing $d_2=g_2-g_3$ and $d_3=g_2+g_3$, resulting in 
$$
{\cal F}_{sg}=
 {\textstyle{1\over2}}g_2[(\vec\nabla e_2)^2+(\vec\nabla e_3)^2]
+{\textstyle{1\over2}}g_3\{
 {\textstyle{1\over2}}[(e_{2,1})^2\!-\!(e_{3,1})^2] 
$$
$$
+{\textstyle{1\over2}} [ (e_{2,2})^2\!-\!(e_{3,2})^2] 
-                     [(e_{2,3})^2\!-\!(e_{3,3})^2] 
$$
\begin{equation}
+\sqrt{3}(e_{2,1}e_{3,1}\!-\!e_{2,2}e_{3,2}) \} \ .
\end{equation}

We seek the solution linking the variants with four-fold axes in the 1 and 2 
directions, Eqs. (3b) and (3c); 
the results for this pair are simpler than for the others. 
The method is easily extended to treat a twin band. 
We assume a solution with $e_4$, $e_5$, $\omega_1$ and $\omega_2=0$ all 
identically zero, and with the other four strains (and $\omega_3$) 
independent of $x_3$. 
The strains are not independent for physical settings like a twin wall 
where the strains depend on position; 
rather, they are linked by the compatibility relations (necessary and 
sufficient conditions that the strains be derivable from the displacement). 
The nine first-order relations, of the form $u_{i,jk}=u_{i,kj}$, involve the 
first derivatives of the strains and the rotation components $\omega_i$. 
The more familiar second-order relations, which involve the second 
derivatives of the strains, are easily obtained by differentiation 
to eliminate the $\omega_i$. 
By virtue of the above assumptions, only the relations 
\begin{equation}
\begin{array}{rl}
\partial_2\left(\sqrt{2}e_1+\sqrt{3}e_2+e_3\right)/\sqrt{6}
& = \partial_1\left(e_6+\omega_3\right) \ ,\\
\partial_1\left(\sqrt{2}e_1-\sqrt{3}e_2+e_3\right)/\sqrt{6}
& =  \partial_2\left(e_6-\omega_3\right) \ , \\
\partial_1\left(e_1-\sqrt{2}e_3\right) & =  0 \ ,\\
\partial_2\left(e_1-\sqrt{2}e_3\right) & =  0 \ ,
\end{array}
\end{equation}
need be considered. 
We try for functions of $X=x_1 \cos\beta +x_2 \sin\beta$ alone; 
a similar 1D solution is not possible for a cubic-tetragonal soliton. 
The compatibility relations are then 
\begin{eqnarray}
   \sin\beta\left(\sqrt{2}e_1+\sqrt{3}e_2+e_3\right)/\sqrt{6}
&-&\cos\beta\left(e_6+\omega_3\right)=K_1\ ,\nonumber\\
   \cos\beta\left(\sqrt{2}e_1-\sqrt{3}e_2+e_3\right)/\sqrt{6}
&-&\sin\beta\left(e_6-\omega_3\right)=K_2\ ,\nonumber\\
e_1&-&\sqrt{2}e_3=K_3\ .         
\end{eqnarray}
The constants $K_1$, $K_2$ and $K_3$ are evaluated from the boundary 
conditions at $X=\pm\infty$, namely 
$e_2=\pm{1\over2}\sqrt{3}e_{30}$, $e_3=-{1\over2}e_{30}$, $e_1=e_6=0$, 
and $\omega_3=\pm\Omega$. 
One finds easily that a solution is possible only if $\cos 2\beta=0$, and 
so $X=(x_1\pm x_2)/\sqrt{2}$; 
that is, the walls lie in the $110$ or $1\bar{1}0$ planes. 
In this way, we find the key new results relating $e_1$ and $e_6$ 
to the deviatoric strains:  
\begin{eqnarray} 
e_1(X)&=&\sqrt{2}\left[e_3(X)+e_{30}/2\right]              \ , \nonumber\\
e_6(X)&=&\pm\sqrt{3/2} \left[e_3(X)+e_{30}/2\right]        \ , 
\end{eqnarray} 
plus $\omega_3(X)=\pm e_2(X)/\sqrt{2}$. 
These results are independent of the details of the free-energy density 
(they apply whether or not the dilatational and 
shear energies appear in ${\cal F}$). 

Since the compatibility relations are satisfied, we can use the density 
\begin{eqnarray} 
{\cal F}_{twin}=
& & \frac{A_{ds}}{2}\left(e_3+\frac{e_{30}}{2}\right)^2
   +\frac{A_2   }{2}\left(e_2^2+e_3^2         \right)
   -\frac{B_2   }{3}\left(e_3^3-3e_2^2e_3     \right)    \nonumber\\
& &+\frac{C_2   }{4}\left(e_2^2+e_3^2         \right)^2
   +\frac{D_2   }{2}\left(\frac{de_2}{dX}     \right)^2
   +\frac{D_3   }{2}\left(\frac{de_3}{dX}     \right)^2
\end{eqnarray} 
and minimize $F$ with respect to $e_2$ and $e_3$. 
Here $A_{ds}=2A_1+{3\over2}A_4$, 
$D_2={1\over4}d_2+{3\over4}d_3$, and $D_3={3\over4}d_2+{1\over4}d_3$; 
adding the diagonal invariants ${1\over2}d_1(\vec\nabla e_1)^2$ and 
${1\over2}d_4[(\vec\nabla e_4)^2+(\vec\nabla e_5)^2+(\vec\nabla e_6)^2]$ 
(with $d_1$ and $d_4$ $>0$) to the density of Eq. (5) leaves 
$D_2$ unchanged while adding $2d_1+{3\over2}d_4$ to $D_3$. 
The corresponding Euler-Lagrange equations are 
\renewcommand{\theequation}{12\alph{equation}}
\setcounter{equation}{0}
\begin{equation}
A_2e_2+2B_2e_2e_3+C_2e_2(e_2^2+e_3^2)=D_2d^2e_2/dX^2\ ,
\end{equation}
\begin{eqnarray}
&& A_{ds}(e_3+e_{30}/2)+A_2e_3+B_2(e_2^2-e_3^2)+C_2e_3(e_2^2+e_3^2)
\nonumber \\
 && \hspace{.3in} =D_3d^2e_3/dX^2 \ .
\label{12b}
\end{eqnarray}
\renewcommand{\theequation}{\arabic{equation}}
\setcounter{equation}{12}
The term $A_{ds}(e_3+e_{30}/2)$ in Eq. (\ref{12b}), new with this article, 
results from satisfying the compatibility relations.
The same equations are obtained, after integrations, on using the density of 
Eqs.(2) plus (3), demanding that $F$ be stationary with respect to the 
displacement, and only then using Eqs.(10). 
The boundary conditions are 
\begin{equation} 
    e_2(\pm\infty)=\pm\sqrt{3}e_{30}/2\ \ ,
\ \ e_3(\pm\infty)=-e_{30}/2\ .
\label{14}
\end{equation} 

At $A_2=-2B_2^2/C_2$, which defines the temperature $T_{BK}$, 
the solutions are\cite{bk84}  
\begin{equation} 
    e_2=(\sqrt{3}e_{30}/2)\tanh(\kappa X)\ \ , \ \ e_3=-e_{30}/2
\end{equation} 
with $\kappa^2=3B_2^2/(2C_2D_2)$; 
the $A_{ds}$ term and the dilatational and shear strains all vanish 
identically. 
We note that $T_{BK}$ may possibly be identified experimentally as the 
temperature where $e_{30}$ ($=2B_2/C_2$) is three times the value at $T_c$. 

At other temperatures, the strains $e_1$ and $e_6$ are not zero, and 
the equations must be solved numerically. 
We define the reduced temperature $\tau=(T-T_0)/(T_c-T_0)$; 
then $\tau=1$ at the transition and $\tau=-9$ at $T=T_{BK}$. 
We also take $D_3=D_2$. 

\begin{figure}[ht]
\epsfysize=2.9in
\epsfbox[60 240 570 670]{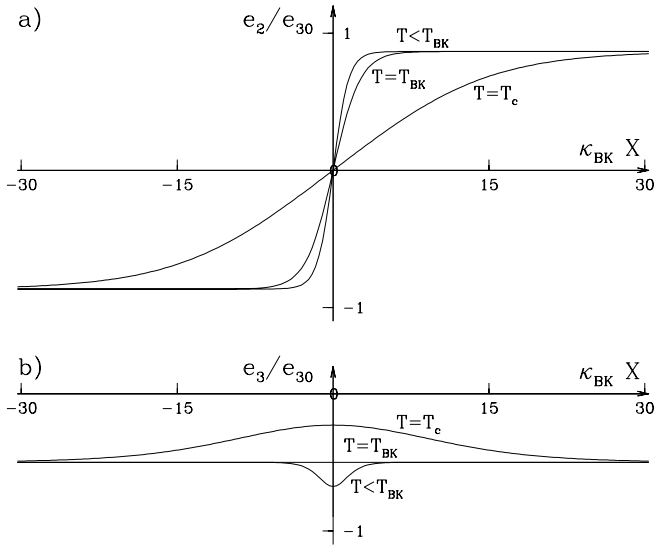}
\caption{
Order parameters $e_2$ and $e_3$ as functions of position.  The
horizontal axis  is scaled by
$\kappa$ (see Eq. 14) defined at $T_{BK}$.
The parameter $A_{ds}$ is $A_2^\prime(T_c-T_0)$; 
the temperatures $T=T_c$, $T=T_{BK}$ and $T<T_{BK}$ correspond to
$\tau=1$, $-9$ and $-50$ respectively.
}
\end{figure}

Figure 1 shows $e_2$ and $e_3$ as functions of $X$ at three different 
temperatures, $T_c$ (which is $>T_{BK}$), $T_{BK}$ and $T<T_{BK}$, 
as determined from numerical solution of Eqs. (12) and (13).
The value $A_{ds}=A_2^\prime(T_c-T_0)=A_2(T=T_c)$, which is rather soft, 
was chosen for display purposes; for larger values, $e_3$ remains close 
to $-e_{30}/2$.
One sees that $e_3+{1\over2}e_{30}$ is $>0$ or $<0$ for $T>T_{BK}$ or 
$<T_{BK}$ respectively. 
From Eqs. (10) then, the wall region is dilated near $T_c$ and compressed 
below $T_{BK}$. 

To estimate the size of these effects, we use data for FePd alloys.
The data quoted in 
Ref.\onlinecite{hongolson} give $T_c=268.6$K, $T_0=265$K and $T_{BK}=233$K. 
Combining these with data from Refs. \onlinecite{muto}, we find that 
$A_{ds}/A_2(T=T_c)$ has a lower limit of $\approx 200$ based on the
smallest observed values for $A_2$; we use a more conservative
estimate of 400.  
Between $T_c$ and $T_{BK}$, the volume change $\Delta V/V$ at the centre of the 
wall reaches a maximum at $T\approx 258$ K, with a very 
small value $\approx 10^{-4}$.  At $T=0$ we find 
$\Delta V/V \approx -10^{-2}$. 

\begin{figure}[ht]
\epsfysize=3.3in
\epsfbox[50 180 560 690]{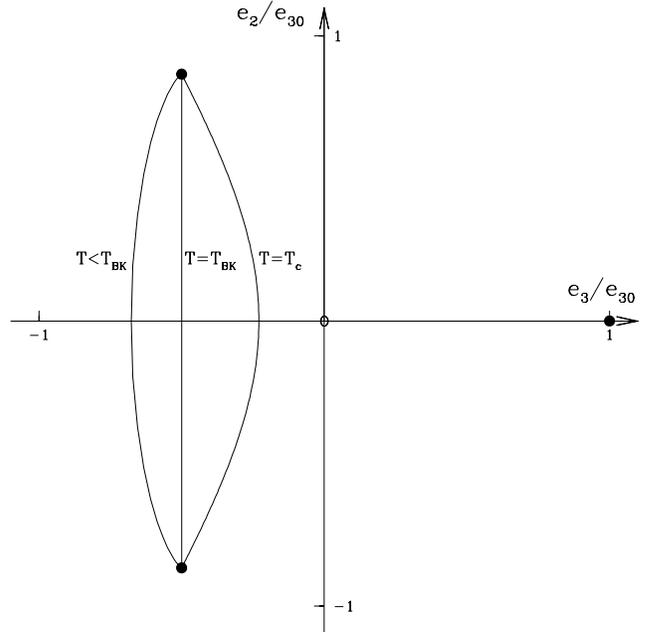}
\caption{
Trajectories in the ($e_3,e_2$) plane for a twin wall linking two of the
three tetragonal variants (solid circles).
The parameter $A_{ds}$ and the temperatures are as in Figure 1.
}
\end{figure}

Figure 2 shows twin-wall trajectories in the $(e_3,e_2)$ plane. 
The trajectories bow toward the third variant for $T_{BK}<T<T_c$, and away 
for $T<T_{BK}$. 
They shift toward the vertical for larger $A_{ds}$.

Reference\onlinecite{yamada}, in effect, assumed that $A_1=A_6=0$ and so the 
term $A_{ds}(e_3+e_{30}/2)$ was absent from their differential equations. 
Solution of these equations, done only near $T_c$, gave trajectories which 
passed close to the origin. 
We point out however that the origin is not the cubic state but rather a 
highly dilated, highly sheared state with $e_1=e_{30}/\sqrt{2}$ and 
$e_6=\pm \sqrt{3/8}e_{30}$, as seen from Eqs.(10). 
Reference\onlinecite{hongolson}, which also assumed in effect that 
$A_1=A_6=0$, proposed a trajectory that is a $2\pi/3$ circular arc centered 
at the origin.
As evident from Figure 2, this is possibly useful only for $T\ll T_{BK}$.

The wall energy $W$ (per unit area) is the energy required to form an 
interface between two variants: 
\begin{equation}
W=\int_{-\infty}^\infty\left({\cal F}_{twin}-{\cal F}_h\right)\ dX
\end{equation}
where ${\cal F}_{twin}$ is the density (11) for the twin-wall solution of Eqs. 
(12) and (13), and 
\begin{equation}
{\cal F}_h={1\over2}A_2e_{30}^2 -{1\over3}B_2e_{30}^3 +{1\over4}C_2e_{30}^4 
\end{equation}
is the density for a single variant. 
Although not directly observable, 
$W$ has physical content and so we provide the 
following.

\begin{figure}[ht]
\epsfysize=3.1in
\epsfbox[20 180 570 690]{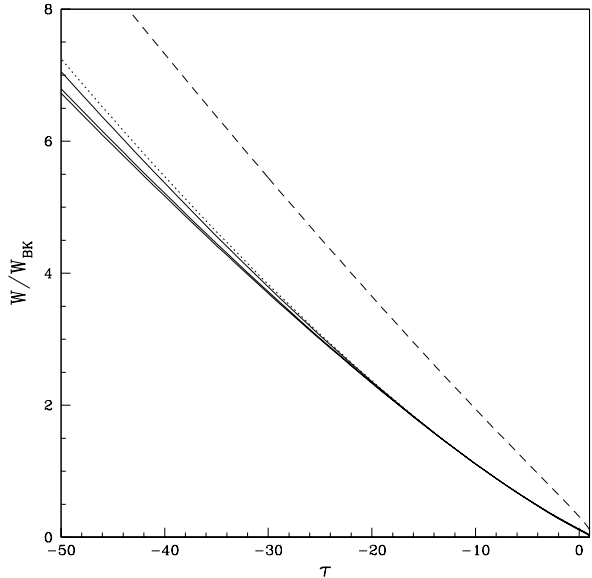}
\caption{
Numerical and variational wall energies as functions of the
dimensionless temperature $\tau$.  The vertical axis is scaled by the wall
energy at $T_{BK}$.
From lower to upper, the solid curves are results from the solution of the
differential equations for $A_{ds}=(1, 10, 100)\times A_2^\prime(T_c-T_o)$.
The short-dashed and long-dashed curves are the bounds of Eqs. (17) and (18) 
respectively. 
}
\end{figure}

Figure 3 shows the wall energy $W$ as a function of temperature, as determined
numerically for three different values of $A_{ds}$; 
it shows also two variational approximations which we now obtain. 

If we consider Eq.(14) as a trial function, with $\kappa$ an adjustable 
parameter, we find 
\begin{equation}
W\leq e_{30}^3\sqrt{3C_2D_2/8} 
\end{equation}
at the optimal $\kappa$, namely $\kappa=e_{30}\sqrt{3C_2/(8D_2)}$; 
the $T$ dependence is in $e_{30}$. 
From Figure 3, this does very well over the temperatures examined, not the 
least because it is an equality at $T_{BK}$ for all $A_{ds}$. 
The coefficients $A_{ds}$ and $D_3$ do not appear because of the form of our 
trial function; 
explicitly, in obtaining Eq.(17), we did {\it not} drop the corresponding 
terms in the density, and so Eq.(17) is valid independent of the magnitude 
of the dilatational and shear energies and strains. 
Equation (17) seems to be an equality also in the limit $A_{ds}\to\infty$ 
for all $T$.

The wall energy was estimated previously\cite{hongolson}, 
using the circular trajectory described above.
On setting $A_1=A_6=A_{ds}=0$, taking $D_3=D_2$, and using\cite{hongolson} 
$e_3=e_{30}\cos\phi$, $e_2=e_{30}\sin\phi$, with $\frac{2\pi}{3}\leq\phi 
\leq\frac{4\pi}{3}$, we find 
\begin{equation}
W\leq\sqrt{64 B_2 D_2 e_{30}^5 /27}\ .
\label{18}
\end{equation}
Unlike Eq.(17), this assumes that $A_1=A_6=0$; 
including dilatation and shear only increases the value relative to Eq.(17). 
Figure 3 shows that Eq.(\ref{18}) is considerably poorer than Eq.(17) for the 
temperatures examined; it is a factor of 3 larger at $T_c$. 
But the trajectories bend toward the Hong-Olson arc at lower $T$, and so 
Eq.(\ref{18}) increases with decreasing $T$ less strongly than Eq.(17) 
(as $e_{30}^{5/2}$ rather than as $e_{30}^3$). 
Indeed, for $\tau<-151$, Eq.(\ref{18}) is actually better than Eq.(17); 
but this temperature may not be accessible since the strain there is 
$\approx 10$ times that at $T_c$. 
For Fe$_{70}$Pd$_{30}$, $\tau =-151$ is inaccessible ($T<0$).

In summary, we have found for the first time the solution for the C-T twin 
wall, at all $T$, in the physical case with dilatational and shear energies 
and strains. 
The dilatation and shear strains, which are localized near the interface, 
  change sign at $T_{BK}$. The magnitudes (at most 1\% in Fe-Pd alloys) may 
  however be too small to detect.

\acknowledgments
This research was supported by the Natural Sciences and Engineering 
Research Council (Canada).

\end{document}